\documentclass[12pt]{article}
\usepackage{fontspec}
\usepackage[protrusion=true]{microtype}
\usepackage{geometry}
\usepackage{graphicx}
\usepackage{amsmath}
\usepackage{booktabs}
\usepackage{siunitx}
\usepackage[dvipsnames]{xcolor}
\usepackage[hidelinks]{hyperref}
\usepackage{url}
\usepackage{setspace}
\usepackage[font=small,labelfont=bf,skip=6pt]{caption}
\usepackage{titlesec}
\titlespacing*{\section}{0pt}{1.2em}{0.6em}
\titlespacing*{\subsection}{0pt}{1em}{0.5em}
\titlespacing*{\subsubsection}{0pt}{0.9em}{0.4em}
\usepackage{enumitem}

\usepackage{pgfplots}
\pgfplotsset{compat=1.18}
\usepgfplotslibrary{fillbetween}
\usetikzlibrary{patterns}
\definecolor{capblue}{HTML}{2C6FB5}
\definecolor{capred}{HTML}{C0392B}
\definecolor{capgray}{HTML}{7F8C8D}
\definecolor{capgreen}{HTML}{2E8B57}

\usepackage{unicode-math}
\setlist{leftmargin=2em, itemsep=0.25em, topsep=0.5em, partopsep=0pt}
\newcommand{\pD}{p_D}
\newcommand{\PDx}{p(D \mid x)}
\newcommand{\slope}{\mathrm{slope}}
\newcommand{\AR}{\mathrm{AR}}
\newcommand{\WOE}{\mathrm{WOE}}
\newcommand{\IV}{\mathrm{IV}}
\newcommand{\E}{\mathbb{E}}

\begin{document}
\onehalfspacing

\title{The Gini--Bayes Connection:\\[0.3em]
\large The CAP Slope as Bayes' Theorem, with Applications to\\
Weight of Evidence, Somers' \emph{D}, and Calibration}
\author{Denis Burakov}
\date{June 2026}
\maketitle

\begin{abstract}
\noindent
The probabilistic reading of the cumulative accuracy profile (CAP) has a long industry
lineage. Falkenstein, Boral and Carty (2000) state, in discrete form, that the default
rate at a score percentile equals the portfolio average rate times the local slope of
the power curve; van der Burgt (2008, 2019) formalizes this as the continuous identity
$\PDx = \pD\, dy/dx$ and imports the continuous form as a working fact; Tasche (2009) analyzes the
resulting calibration method; Voloshyn and Voloshyn (2023) substitute Bayes' theorem,
$f(x\mid D)=\PDx f(x)/\pD$, into the area integral and write the Gini as a functional of the
calibration curve. The slope itself is already in the lineage (van der Burgt's $dy/dx$ is the
ratio of the two cumulative differentials), but it enters as a cited working fact, never as
Bayes' theorem. We make that identification explicit and draw out its consequences. First, the
CAP slope \emph{is} Bayes' theorem in cumulative coordinates: the standardized PD it recovers is
the posterior probability rescaled by the prior. The weight of the paper then falls on two results this
reading unlocks. The odds form places the weight of evidence (the log of the likelihood ratio, i.e.\ the
Bayes factor) and the information value inside one geometry (the weight of evidence at a point
is the log of the ratio of the ``bad'' and ``good'' CAP slopes). The accuracy ratio, Somers' $D_{xy}$,
and the Gini $(2A-1)/(1-\pD)$ are revealed as one number computed three ways. Run in \emph{comparison
mode} (realized outcomes against model claims), the same identity recovers the reliability
diagram in cumulative coordinates, with the sign of the gap between the empirical and
model-implied Gini coefficients as a calibration diagnostic. A worked five-band example carries every
identity in discrete form, and a kernel-density example extends them to the continuous case.
\end{abstract}

\section{Introduction}

Discriminatory power and calibration are the two axes along which a credit rating system is
judged. Power asks whether the model orders borrowers correctly; calibration asks whether
the probabilities it attaches to them are credible (if we say 5\% probability of default,
it means 5\% of the borrowers will actually default). The cumulative accuracy profile (CAP) and
its summary, the accuracy ratio (or Gini coefficient), are well studied as the standard instruments for
the first question, and a large industry literature treats them as power-only objects: the area
under the curve uses ranks and discards levels.

Yet the separation is not as clean as it looks. The probabilistic reading of the CAP has an
industry lineage that quietly carries calibration inside a power object. Falkenstein, Boral
and Carty (2000), in the appendix of the Moody's RiskCalc methodology, state a discrete
relationship between the default-frequency curve and the power curve: the default rate at a
score percentile equals the portfolio mean rate times the local rise of the power curve.
Van der Burgt (2008) operationalizes this for low-default portfolios by fitting the CAP to a
one-parameter concave family and differentiating it to read off per-grade PDs; Tasche (2009)
analyzes that approach; van der Burgt (2019) writes the relationship as the continuous slope
identity
\begin{equation}\label{eq:slope-id}
\PDx \;=\; \pD\,\frac{dy}{dx},
\end{equation}
attributes it to Falkenstein et al.\ (2000), and uses the same geometry to construct the
rating scale itself. Voloshyn and Voloshyn (2023) substitute Bayes' theorem into the area
integral and obtain the Gini as a functional of the calibration curve.

In every step of this chain the slope \eqref{eq:slope-id} enters by citation. Van der Burgt's
$dy/dx$ is, in CAP coordinates, already the ratio of the two cumulative differentials; what is
never written is that this ratio is simply Bayes' theorem
\begin{equation}\label{eq:bayes-intro}
f(x\mid D)\;=\;\frac{\PDx\,f(x)}{\pD}
\end{equation}
in cumulative coordinates: $dF(x\mid D)/dF(x) = f(x\mid D)/f(x) = \PDx/\pD$, the ratio of the
differential of the numerator to the differential of $F(x)$. We make this identification
explicit; the weight of the paper falls on what it unlocks, the odds-form scorecard geometry and
the comparison-mode calibration diagnostic.

\paragraph{Contributions.} Our contribution is threefold.
\begin{enumerate}
\item \textbf{We make the identity explicit.} Van der Burgt's slope $dy/dx$ is, in CAP
coordinates, exactly $dF(x\mid D)/dF(x)$; reading it through Bayes shows the CAP slope \emph{is}
the standardized PD $\PDx/\pD$, the posterior rescaled by the prior (Section~\ref{sec:slope}).
This re-reads a known quantity rather than proposing a new estimator; its value is what it
unlocks below.
\item \textbf{We extend it through the odds form.} The ratio of the bad and good CAP slopes
is the Bayes factor in favor of default, its logarithm is the weight of evidence, and the
Jeffreys divergence of the two class-conditional densities is the information value
(Section~\ref{sec:woe}). The weight of evidence is recoverable from the slope profile alone,
while the information value additionally needs the band counts.
\item \textbf{We run the identity in comparison mode.} Feeding it realized outcomes and
model claims produces an empirical and a model-implied CAP whose slope-by-slope agreement,
together with the level condition $\hat p_D = \pD$, is exactly calibration
(Section~\ref{sec:calib}): the reliability diagram in cumulative coordinates, with the gap
between the two Gini coefficients as its one-number summary. Along the way we show that the accuracy
ratio, Somers' $D_{xy}$, and the Gini are the same number computed three ways
(Section~\ref{sec:somers}).
\end{enumerate}

We take the rating bands as given. Their construction is itself an application of this
geometry (van der Burgt 2019), and the regulatory requirement of a monotone PD scale is
precisely the concavity condition: slopes decreasing along the sort order. We treat the
discrete case throughout for clean derivations; Section~\ref{sec:continuous} gives the
continuous (density-estimation) reading.

One sentence summarizes the stance of the paper:
\begin{quote}
\emph{Slope $\times$ level $=$ PD is identity; the shape of the slope profile is assumption.}
\end{quote}
\begin{equation}
\underbrace{\text{PD}(x)}_{\PDx}
\;=\;
\underbrace{\langle D\rangle}_{\pD}
\,\cdot\,
\underbrace{\frac{dy}{dx}}_{\slope(x)},
\qquad \frac{dy}{dx}=\frac{dF(x\mid D)}{dF(x)}.
\end{equation}

\section{Setup and definitions}\label{sec:setup}

Borrowers carry an ordinal rating grade $x$, sorted from worst (riskiest) to best credit quality. The
default indicator is $Y\in\{0,1\}$, with $D=\{Y=1\}$. Three distributions organize
everything that follows:
\begin{itemize}
\item $f(x)$, the \textbf{marginal} (population) probability of grade $x$;
\item $\PDx$, the \textbf{posterior} conditional probability of default at grade $x$
(the calibration curve, in business terms the default rate);
\item $f(x\mid D)$, the \textbf{likelihood}: how grade $x$ is distributed among defaulters
(the class-conditional density of $x$ given default).
\end{itemize}
The unconditional (prior) default rate is the $f$-weighted average of the calibration curve,
\begin{equation}\label{eq:prior}
\pD \;=\; \int_0^1 \PDx\,f(x)\,dx
\qquad\text{(discretely, }\textstyle\pD=\sum_x \PDx f(x)\text{)}.
\end{equation}
Bayes' theorem ties the three together, the form used by Voloshyn and Voloshyn (2023):
\begin{equation}\label{eq:bayes}
\underbrace{f(x\mid D)}_{\text{likelihood}}
=\frac{\overbrace{\PDx}^{\text{posterior}}\,\overbrace{f(x)}^{\text{marginal}}}
{\underbrace{\pD}_{\text{prior}}}.
\end{equation}
Read in words, $f(x\mid D)$ answers: \emph{if I pick a defaulter at random, what is the
probability they came from grade $x$?} A grade contributes more defaulters when its default rate
$\PDx$ is high (risky) and when its population share $f(x)$ is large (many borrowers
sit there); the denominator $\pD$ normalizes the shares to sum to one.

\paragraph{Running example.} Table~\ref{tab:data} is the five-band portfolio we carry through
the paper. It has $N=115$ borrowers, $B=20$ defaulters, and prior $\pD = 20/115 = 0.1739$.

\begin{table}[htbp]
\centering
\small
\begin{tabular}{l S[table-format=3.0] S[table-format=2.0] S[table-format=2.0]
  S[table-format=1.4] S[table-format=1.4] S[table-format=1.4] S[table-format=1.4]}
\toprule
Band & {Count} & {Goods} & {Bads} & {$\PDx$} & {$f(x)$} & {$f(x\mid D)$} & {$f(x\mid\bar D)$}\\
\midrule
A & 24 & 23 & 1 & 0.0417 & 0.2087 & 0.0500 & 0.2421 \\
B & 36 & 32 & 4 & 0.1111 & 0.3130 & 0.2000 & 0.3368 \\
C & 25 & 20 & 5 & 0.2000 & 0.2174 & 0.2500 & 0.2105 \\
D & 20 & 15 & 5 & 0.2500 & 0.1739 & 0.2500 & 0.1579 \\
E & 10 &  5 & 5 & 0.5000 & 0.0870 & 0.2500 & 0.0526 \\
\midrule
Total & 115 & 95 & 20 & & & & \\
\bottomrule
\end{tabular}
\caption{The five-band running example. $f(x)=\text{count}/115$, $f(x\mid D)=\text{bads}/20$,
$f(x\mid\bar D)=\text{goods}/95$. Prior $\pD=0.1739$.}
\label{tab:data}
\end{table}

\begin{figure}[htbp]
\centering
\begin{tikzpicture}
\begin{axis}[
  width=0.5\textwidth, height=5.2cm,
  ybar, bar width=7pt,
  ymin=0, ymax=0.58,
  symbolic x coords={A,B,C,D,E}, xtick=data,
  ylabel={probability}, xlabel={band $x$},
  legend style={at={(0.5,1.02)},anchor=south,draw=none,legend columns=2,font=\footnotesize},
  tick label style={font=\footnotesize}, label style={font=\footnotesize},
  enlarge x limits=0.15,
]
\addplot[fill=capred!75,draw=capred] coordinates {(A,0.0417)(B,0.1111)(C,0.2000)(D,0.2500)(E,0.5000)};
\addplot[fill=capgray!55,draw=capgray] coordinates {(A,0.2087)(B,0.3130)(C,0.2174)(D,0.1739)(E,0.0870)};
\legend{$\PDx$ (default rate),$f(x)$ (share)}
\end{axis}
\end{tikzpicture}\hfill
\begin{tikzpicture}
\begin{axis}[
  width=0.5\textwidth, height=5.2cm,
  ybar, bar width=7pt,
  ymin=0, ymax=0.35,
  symbolic x coords={A,B,C,D,E}, xtick=data,
  ylabel={probability}, xlabel={band $x$},
  legend style={at={(0.5,1.02)},anchor=south,draw=none,legend columns=2,font=\footnotesize},
  tick label style={font=\footnotesize}, label style={font=\footnotesize},
  enlarge x limits=0.15,
]
\addplot[fill=capgray!55,draw=capgray] coordinates {(A,0.2087)(B,0.3130)(C,0.2174)(D,0.1739)(E,0.0870)};
\addplot[fill=capblue!75,draw=capblue] coordinates {(A,0.0500)(B,0.2000)(C,0.2500)(D,0.2500)(E,0.2500)};
\legend{$f(x)$ (population),$f(x\mid D)$ (defaulters)}
\end{axis}
\end{tikzpicture}
\caption{\textbf{Left:} the two ingredients of Bayes: the calibration curve $\PDx$ and the
population share $f(x)$. \textbf{Right:} Bayes reweights the population $f(x)$ (grey) into the
defaulter mix $f(x\mid D)$ (blue), pulling mass toward the riskier bands. Grey exceeds blue in
the safe bands A--B, where defaulters are under-represented relative to the population, and blue
exceeds grey in the risky bands C--E, where they are over-represented; the crossover is the
reweighting by the default rate $\PDx$. The denominator $\pD$ normalizes $f(x\mid D)$ to sum to one.}
\label{fig:bars}
\end{figure}
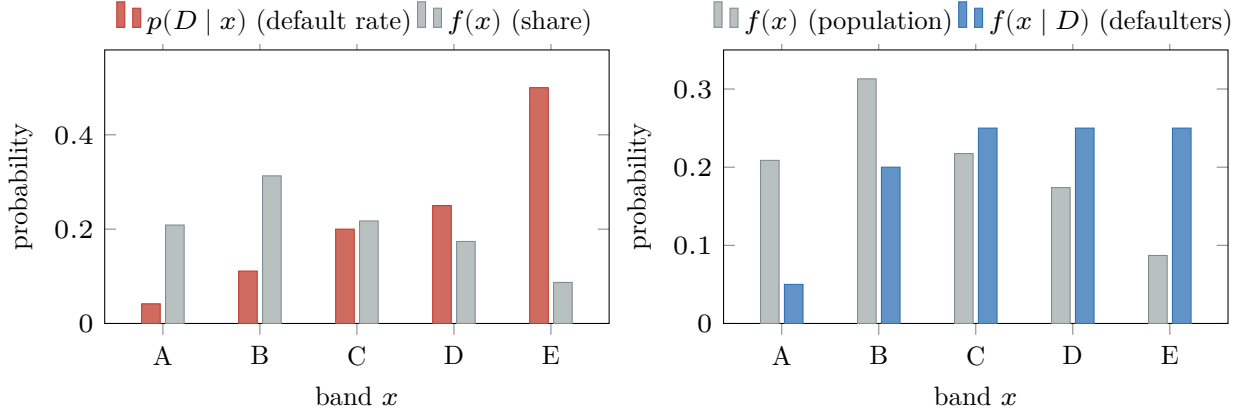

A one-line check of Bayes \eqref{eq:bayes} on band E (writing $x{=}E$ for the rating):
\begin{equation}
f(x{=}E\mid D)=\frac{p(D\mid x{=}E)\,f(x{=}E)}{\pD}
=\frac{0.50\times \tfrac{10}{115}}{\tfrac{20}{115}}
=0.50\times\frac{10}{20}=0.25,
\end{equation}
matching $5$ defaulters out of $20$. Band E has the highest default rate ($50\%$) but the
smallest population share; band B has a low default rate ($11\%$) but the largest share. Bayes
balances the two into the defaulter mix $f(x\mid D)$. Figure~\ref{fig:bars} shows the inputs
and the Bayes reweighting side by side.

\section{The CAP slope is Bayes' theorem}\label{sec:slope}

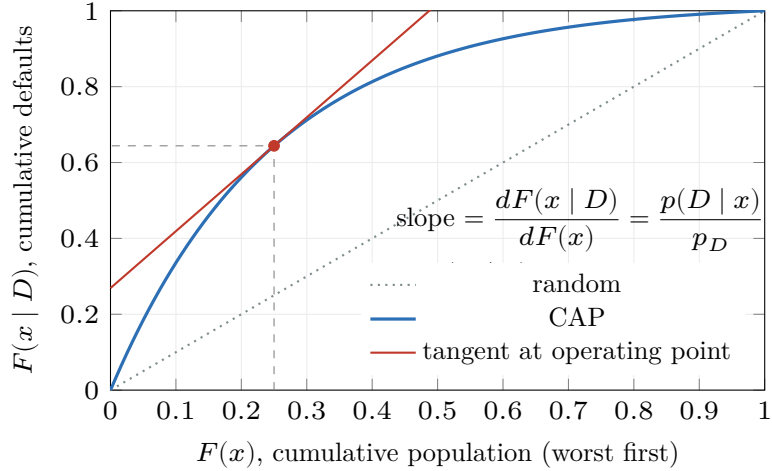
\begin{figure}[htbp]
\centering
\begin{tikzpicture}
\begin{axis}[
  width=0.62\textwidth, height=6.6cm, xmin=0,xmax=1,ymin=0,ymax=1.0,
  xlabel={$F(x)$, cumulative population (worst first)},
  ylabel={$F(x\mid D)$, cumulative defaults},
  tick label style={font=\footnotesize}, label style={font=\footnotesize},
  grid=major, grid style={gray!15}, clip=true,
  legend style={at={(0.97,0.04)},anchor=south east,draw=none,font=\footnotesize},
]
\addplot[capgray,dotted,thick] coordinates {(0,0)(1,1)}; \addlegendentry{random}
\addplot[capblue,very thick,domain=0:1,samples=120] {(1-exp(-4*x))/(1-exp(-4))};
\addlegendentry{CAP}
\addplot[capred,thick,domain=0:0.49,samples=2] {0.644 + 1.50*(x-0.25)};
\addlegendentry{tangent at operating point}
\draw[gray,dashed] (axis cs:0.25,0) -- (axis cs:0.25,0.644) -- (axis cs:0,0.644);
\addplot[only marks,capred,mark=*,mark size=2pt] coordinates {(0.25,0.644)};
\node[anchor=west,font=\footnotesize,align=left] at (axis cs:0.42,0.40)
 {slope $=\dfrac{dF(x\mid D)}{dF(x)}=\dfrac{\PDx}{\pD}$\\[3pt]
  so $\PDx=\pD\cdot\text{slope}$};
\node[font=\footnotesize,anchor=north] at (axis cs:0.25,-0.015) {$x_0$};
\end{axis}
\end{tikzpicture}
\caption{The standardized PD as the tangent slope of the CAP at an operating point $x_0$
(here a smooth CAP, van der Burgt's exponential family). The slope is $\PDx/\pD$; multiplied by
$\pD$ it returns the per-grade PD. Above the $45^\circ$ line the slope exceeds one (riskier than
average); below it the slope is less than one (safer). This is the differential reading van der
Burgt (2019) uses, here identified as Bayes' theorem in cumulative coordinates.}
\label{fig:tangent}
\end{figure}

The CAP curve plots the cumulative share of default risk captured, $F(x\mid D)$, against the
cumulative share of population, $F(x)$, both accumulated worst-first: $F(x\mid D)$ is the
running share of defaulters down to score $x$ and $F(x)$ the running share of all borrowers, so
their densities are $f(x\mid D)$ and $f(x)$. Bayes \eqref{eq:bayes} expresses the defaulter
density through the calibration curve. The \emph{slope} of the CAP at an operating point is the
derivative of one cumulative against the other, and by the chain rule it is the ratio of the
two densities:
\begin{equation}\label{eq:cap-slope}
\boxed{\;
\frac{dF(x\mid D)}{dF(x)}
=\frac{f(x\mid D)}{f(x)}
=\frac{\PDx}{\pD}
\;}
\end{equation}
where the last step is exactly Bayes' theorem \eqref{eq:bayes} with the common factor $f(x)$
cancelled. In words: over a thin slice of the population at score $x$, the CAP rises by the
share of defaulters in that slice and runs by the share of borrowers; their ratio is the slope,
and Bayes says that ratio is the standardized default rate. Multiplying back by the prior
recovers the calibration curve,
\begin{equation}\label{eq:pd-from-slope}
\PDx \;=\; \pD\cdot\frac{dF(x\mid D)}{dF(x)},
\end{equation}
which is the slope identity \eqref{eq:slope-id}. The derivation is one line: the CAP slope is
Bayes' theorem in cumulative coordinates, and the quantity it returns,
\begin{equation}
\mathrm{PD}_{\text{st}}(x)\;:=\;\frac{\PDx}{\pD}\;=\;\frac{dF(x\mid D)}{dF(x)},
\end{equation}
is the \textbf{standardized PD}, the posterior rescaled by the prior, i.e.\ how many times
the portfolio-average default risk a borrower at grade $x$ carries. Geometrically the standardized PD is
the slope of the tangent to the CAP at an operating point (Figure~\ref{fig:tangent}): the curve
tilts up past the $45^\circ$ random line where risk exceeds the average and flattens below it
where risk is low, recovering the per-grade PD as $\PDx=\pD\cdot\text{slope}$, the reading van
der Burgt (2019, EuroBanking presentation, slide~3) uses to turn the CAP into a calibration tool.

\paragraph{Provenance.} Equation~\eqref{eq:pd-from-slope} appears in Falkenstein et al.\ (2000,
App.~4A) in discrete, verbal form: the default frequency at a score percentile equals the mean
default rate times the discrete rise of the power curve, i.e.\ PD $=$ mean-PD $\times$ discrete
CAP slope.\footnote{In their notation, with $\mathrm{prob}(q)$ the default frequency at
percentile $q$ and $\mathrm{power}(q)$ the power-curve height,
$\mathrm{prob}(q)=100\cdot\overline{\mathrm{prob}}\cdot\big(\mathrm{power}(q)-\mathrm{power}(q-1)\big)$,
the $1/100$ being the percentile step.} They neither write it as a derivative nor invoke Bayes. Van der Burgt (2019, Eq.~2.3) gives the
continuous form $\PDx=\pD\,[dy/dx]$ and the per-grade version (his Eq.~2.11)
$P_r=\frac{y_r-y_{r-1}}{x_r-x_{r-1}}\pD$, deriving the latter from Bayes' law (his Eq.~2.10)
applied to a single band. What \eqref{eq:cap-slope} adds is the observation that the \emph{continuous}
identity is itself nothing but Bayes \eqref{eq:bayes} differentiated (the ratio of
$dF(x\mid D)=f(x\mid D)\,dx$ to $dF(x)=f(x)\,dx$), so the per-band Bayes step and the
slope identity are the same statement.

\begin{figure}[htbp]
\centering
\begin{tikzpicture}
\begin{axis}[
  width=0.62\textwidth, height=7cm,
  xmin=0,xmax=1,ymin=0,ymax=1,
  xlabel={$F(x)$, cumulative population share (worst first)},
  ylabel={$F(x\mid D)$, defaults captured},
  legend style={at={(0.98,0.02)},anchor=south east,draw=none,font=\footnotesize},
  tick label style={font=\footnotesize}, label style={font=\footnotesize},
  grid=major, grid style={gray!18},
]
\addplot[capgreen,dashed,thick] coordinates {(0,0)(0.1739,1)(1,1)};
\addlegendentry{perfect model}
\addplot[capgray,dotted,thick] coordinates {(0,0)(1,1)};
\addlegendentry{random}
\addplot[capblue,very thick,mark=*,mark size=1.6pt]
  coordinates {(0,0)(0.0870,0.25)(0.2609,0.50)(0.4783,0.75)(0.7913,0.95)(1,1)};
\addlegendentry{CAP (this portfolio)}
\end{axis}
\end{tikzpicture}
\caption{The CAP curve of Table~\ref{tab:cap}. The slope of each segment is the standardized PD
of that band; multiplied by $\pD$ it returns the band's default rate
(Eq.~\ref{eq:pd-from-slope}). The perfect-model boundary reaches $(\pD,1)=(0.1739,1)$, not
$(0,1)$, the fact behind the $1-\pD$ correction in Section~\ref{sec:somers}.}
\label{fig:cap}
\end{figure}
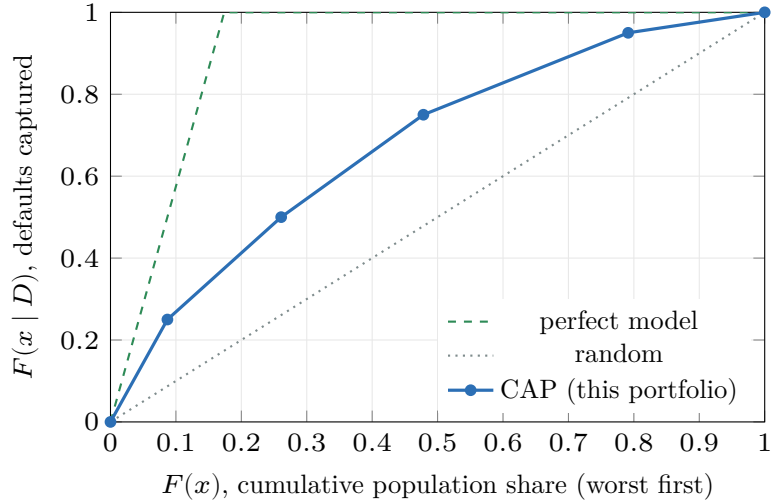

\paragraph{The CAP of the running example.} Accumulating Table~\ref{tab:data} worst-first
(E, D, C, B, A) gives the curve in Figure~\ref{fig:cap}.

\begin{table}[htbp]
\centering
\small
\begin{tabular}{l S[table-format=1.4] S[table-format=1.4] S[table-format=1.4] S[table-format=1.4]}
\toprule
Sorted (worst first) & {$f(x)$} & {$F(x)$} & {$f(x\mid D)$} & {$F(x\mid D)$}\\
\midrule
E & 0.0870 & 0.0870 & 0.2500 & 0.2500 \\
D & 0.1739 & 0.2609 & 0.2500 & 0.5000 \\
C & 0.2174 & 0.4783 & 0.2500 & 0.7500 \\
B & 0.3130 & 0.7913 & 0.2000 & 0.9500 \\
A & 0.2087 & 1.0000 & 0.0500 & 1.0000 \\
\bottomrule
\end{tabular}
\caption{Cumulative coordinates of the CAP curve, sorted worst-first.}
\label{tab:cap}
\end{table}

\section{The odds form: weight of evidence, Bayes factor, information value}\label{sec:woe}

Everything above used the ``bad'' CAP. The complementary ``good'' CAP accumulates the non-defaulter
density $f(x\mid\bar D)$, and its slope is the standardized survival rate:
\begin{equation}
\frac{dF(x\mid D)}{dF(x)}=\frac{\PDx}{\pD},
\qquad
\frac{dF(x\mid\bar D)}{dF(x)}=\frac{1-\PDx}{1-\pD}.
\end{equation}
The two slopes are not independent: since $f(x)=\pD f(x\mid D)+(1-\pD)f(x\mid\bar D)$, they
satisfy the mixture identity $\pD\,\slope_{\text{bad}}(x)+(1-\pD)\,\slope_{\text{good}}(x)=1$ at
every point, so the bad CAP rises above the diagonal exactly where the good CAP dips below.

Bayes' theorem in odds form states posterior odds $=$ likelihood ratio $\times$ prior odds.
Dividing the two slopes is exactly that statement, with $f(x)$ canceling:
\begin{equation}\label{eq:woe}
\underbrace{\frac{\PDx/\pD}{(1-\PDx)/(1-\pD)}}_{\text{ratio of CAP slopes}}
=\underbrace{\frac{\PDx}{1-\PDx}}_{\text{posterior odds}}\cdot
\underbrace{\frac{1-\pD}{\pD}}_{1/\text{prior odds}}
=\frac{f(x\mid D)}{f(x\mid\bar D)}
=e^{\WOE(x)} .
\end{equation}
The middle quantity is the \textbf{Bayes factor} in favor of default, $\mathrm{BF}(x)$, and its
logarithm is the \textbf{weight of evidence}, the name Good (1950) gave the log-Bayes-factor:
\begin{equation}\label{eq:woe-def}
\boxed{\;\WOE(x)=\ln\frac{\slope_{\text{bad}}(x)}{\slope_{\text{good}}(x)}
=\ln\frac{f(x\mid D)}{f(x\mid\bar D)}\;}
\end{equation}
(bad-over-good convention; the industry good-over-bad convention flips the sign). The weight
of evidence at a point is the log of the ratio of the two CAP slopes.

\paragraph{Verification on band E.} The bad slope is $0.25/0.087=2.875$ and the good slope is
$0.5/0.826=0.605$, a ratio of $4.75$. Directly,
$f(E\mid D)/f(E\mid\bar D)=0.25/(5/95)=4.75$, so $\WOE(E)=\ln 4.75=1.558$.

\paragraph{Information value.} The slope table is one step from the information value. Weighting
each band's evidence by the gap between the two likelihoods gives
\begin{equation}\label{eq:iv}
\IV=\sum_x \bigl(f(x\mid D)-f(x\mid\bar D)\bigr)\,\WOE(x),
\end{equation}
the Jeffreys (symmetrized Kullback--Leibler) divergence between the defaulter and non-defaulter
score distributions. Every term is nonnegative because the difference and the log-likelihood
ratio always share a sign (Table~\ref{tab:iv}).

\begin{table}[htbp]
\centering
\small
\begin{tabular}{l S[table-format=1.3] S[table-format=1.3]
  S[table-format=-1.3] S[table-format=-1.3] S[table-format=1.3]}
\toprule
Band & {$f(x\mid D)$} & {$f(x\mid\bar D)$} & {Difference} & {$\WOE(x)$} & {Contribution}\\
\midrule
A & 0.050 & 0.242 & -0.192 & -1.577 & 0.303 \\
B & 0.200 & 0.337 & -0.137 & -0.521 & 0.071 \\
C & 0.250 & 0.211 &  0.039 &  0.172 & 0.007 \\
D & 0.250 & 0.158 &  0.092 &  0.460 & 0.042 \\
E & 0.250 & 0.053 &  0.197 &  1.558 & 0.308 \\
\midrule
Total & & & & & 0.731 \\
\bottomrule
\end{tabular}
\caption{Information value of the running example, $\IV=0.731$. The weight of evidence in the
fifth column is $\ln(\slope_{\text{bad}}/\slope_{\text{good}})$ by Eq.~\eqref{eq:woe-def}.}
\label{tab:iv}
\end{table}

The scorecard trio (standardized PD, weight of evidence, information value) lives largely inside
CAP geometry: the standardized PD is one slope and the weight of evidence is the log-ratio of two
slopes. The information value, however, is the Jeffreys divergence between the two
class-conditional densities, not between the slope profiles, and the distinction is real. The
weight of evidence is a ratio of slopes, so the population factor $f(x)$ cancels and it is
recoverable from the slope profile alone. The information value weights each band by the
difference $f(x\mid D)-f(x\mid\bar D)=f(x)\,\bigl(\slope_{\text{bad}}(x)-\slope_{\text{good}}(x)\bigr)$,
in which $f(x)$ does not cancel; without the band counts the slopes alone do not determine it.
This recovers, from the CAP side, the information-theoretic framework of Sudjianto
and Burakov (2025). The weight of evidence reading as a log-Bayes-factor goes back to
Good (1950, 1985); its use as a model-explanation device follows Alvarez-Melis (2019, 2021).

\section{Accuracy ratio, Somers' \texorpdfstring{$D$}{D}, and the Gini}\label{sec:somers}

The area under the CAP, $A=\int_0^1 F(x\mid D)\,dF(x)$, summarizes power in one number. For
the running example the trapezoidal area is $A=0.6815$. A common shortcut reports
$\AR\approx 2A-1=0.363$, but this is valid only when $\pD\approx 0$. The exact accuracy ratio
carries a $1-\pD$ correction,
\begin{equation}\label{eq:ar-exact}
\AR=\frac{2A-1}{1-\pD}=\frac{2(0.6815)-1}{1-0.1739}=\frac{0.363}{0.826}=0.4395,
\end{equation}
because the perfect-model boundary in CAP coordinates reaches $(\pD,1)$, not $(0,1)$
(Figure~\ref{fig:cap}): the maximum attainable area is $1-\pD/2$, not $1$. Van der Burgt
(2019, Eq.~2.6), who attributes it to Tasche (2005, 2010) (the latter the working-paper
version of the arXiv preprint cited here as Tasche 2009), and Voloshyn and Voloshyn
(2023, Eq.~1, after Thomas 2009) both use the exact
form; van der Burgt (2008) uses the approximation $\AR\approx 2A-1$.

\paragraph{Somers' $D$ via pair counting.} The same $0.4395$ arises with no curve at all, by
counting concordant and discordant good--bad pairs. Take one good and one bad borrower; the
pair is \emph{concordant} if the bad sits in a riskier band than the good, \emph{discordant}
if the good sits in a riskier band, and \emph{tied} if they share a band. Somers'
$D_{xy}$ (Somers, 1962),\footnote{Somers' $D$ is asymmetric. The $2\,\mathrm{AUC}-1$ form used here, normalized by the pairs that differ in the outcome (the good--bad pairs), is $D_{xy}$ in the convention of Newson (2002) and the ROC literature; some sources write $D_{yx}$. The explicit denominator below fixes the meaning regardless of subscript order.} the rank correlation of the default outcome $y$ with the rating $x$, is
\begin{equation}\label{eq:somers}
D_{xy}=\frac{P-Q}{n_{\text{good}}\,n_{\text{bad}}},
\end{equation}
with $P$ concordant and $Q$ discordant pairs (the Kendall convention; the symbol $D$ is reserved
for the default event) and all $n_{\text{good}}n_{\text{bad}}$ good--bad pairs (including ties) in
the denominator. These pairs are drawn from the rating-by-outcome contingency table
(Table~\ref{tab:contingency}); Table~\ref{tab:pairs} carries out the count:
$P=1192$, $Q=357$, ties $=351$, total $=95\times20=1900$, so
\begin{equation}
D_{xy}=\frac{1192-357}{1900}=0.4395 .
\end{equation}

\begin{table}[htbp]
\centering
\small
\begin{tabular}{l r r r}
\toprule
Rating $x$ & Goods & Bads & Total \\
\midrule
A & 23 & 1 & 24 \\
B & 32 & 4 & 36 \\
C & 20 & 5 & 25 \\
D & 15 & 5 & 20 \\
E & 5 & 5 & 10 \\
\midrule
Total & 95 & 20 & 115 \\
\bottomrule
\end{tabular}
\caption{The rating-by-outcome contingency table. Pairing each of the $95$ goods with each of
the $20$ bads gives $n_{\text{good}}n_{\text{bad}}=1900$ good--bad pairs; classifying each as
concordant, discordant, or tied by rating (Table~\ref{tab:pairs}) yields Somers' $D_{xy}$.}
\label{tab:contingency}
\end{table}

\begin{table}[htbp]
\centering
\small
\begin{tabular}{l r r}
\toprule
Band of the good & Concordant (bad riskier) & Discordant (good riskier)\\
\midrule
A (23 goods) & $23\times(4{+}5{+}5{+}5)=437$ & $23\times 0 = 0$ \\
B (32 goods) & $32\times(5{+}5{+}5)=480$ & $32\times 1 = 32$ \\
C (20 goods) & $20\times(5{+}5)=200$ & $20\times(4{+}1)=100$ \\
D (15 goods) & $15\times 5 = 75$ & $15\times(5{+}4{+}1)=150$ \\
E (5 goods)  & $5\times 0 = 0$ & $5\times(5{+}5{+}4{+}1)=75$ \\
\midrule
Total & $1192$ & $357$ \\
\bottomrule
\end{tabular}
\caption{Concordant/discordant good--bad pair counts. ``Bad riskier'' counts, for the goods in
a band, all bads in riskier bands; ``good riskier'' counts all bads in safer bands. Ties (same
band) are $1900-1192-357=351$.}
\label{tab:pairs}
\end{table}

\paragraph{One number, three routes.} The accuracy ratio, Somers' $D_{xy}$, and the
Mann--Whitney form of the ROC Gini coincide exactly:
\begin{equation}\label{eq:three-routes}
\AR=\underbrace{\frac{2A-1}{1-\pD}}_{\text{CAP area}}
=\underbrace{\frac{P-Q}{n_{\text{good}}n_{\text{bad}}}}_{\text{Somers' }D_{xy}}
=\underbrace{2\,\mathrm{AUC}-1}_{\text{ROC}}
=0.4395,
\qquad
\mathrm{AUC}=\frac{P+\tfrac12 T}{n_{\text{good}}n_{\text{bad}}}=0.7197 .
\end{equation}
The CAP route integrates the curve and corrects by $1-\pD$; the pair-counting route never
draws a curve; the ROC route counts the same pairs with ties split. All three are the same
rank statistic, and the $1-\pD$ in \eqref{eq:ar-exact} is precisely what reconciles the
curve-area approach with the pair-counting method.

\section{Calibration in cumulative coordinates}\label{sec:calib}

So far the calibration curve $\PDx$ has been the observed default rate, and the CAP has recovered
it through \eqref{eq:pd-from-slope}. But \eqref{eq:cap-slope}--\eqref{eq:pd-from-slope} hold
for \emph{any} function placed in the $\PDx$ slot, provided $\pD$ is its own $f$-weighted
average. Whether that function coincides with observed default frequencies is a separate
question, one of calibration. Running the identity in comparison mode turns
it into a calibration diagnostic.

\paragraph{Two CAPs.} Sort all borrowers worst-first by the model's score and accumulate two
quantities.
\begin{itemize}
\item \textbf{Empirical CAP} (the outcome): accumulate realized defaults $y_i\in\{0,1\}$,
\begin{equation}
F_{\text{emp}}(x\mid D)=\frac{\sum_{i\le x}y_i}{\sum_i y_i}.
\end{equation}
Its slope, rescaled by the realized level, recovers the observed frequencies:
$\slope_{\text{emp}}(x)\cdot\pD=\E[Y\mid x]$.
\item \textbf{Model-implied CAP} (the promise): accumulate the claimed probabilities $\hat p_i$,
treating each borrower as $\hat p_i$ expected defaults,
\begin{equation}
F_{\text{model}}(x\mid D)=\frac{\sum_{i\le x}\hat p_i}{\sum_i \hat p_i}.
\end{equation}
Its slope recovers the claims: $\slope_{\text{model}}(x)\cdot\hat p_D=\hat p(D\mid x)$, where
$\hat p_D=\sum_x f(x)\,\hat p(D\mid x)$ is the model's average claim.
\end{itemize}
The model-implied CAP is the model's forecast of its own CAP, the curve that would
materialize if the claimed PDs were the ground truth.

\paragraph{Calibration as coincidence of the two curves.} Perfect calibration is the statement
that the two CAPs agree everywhere, which splits into a shape condition and a level condition:
\begin{equation}\label{eq:calib-cond}
\hat p(D\mid x)=\E[Y\mid x]\ \ \forall x
\;\;\Longleftrightarrow\;\;
\hat p_D=\pD
\ \text{ and }\
\slope_{\text{model}}(x)=\slope_{\text{emp}}(x)\ \ \forall x .
\end{equation}
Differentiating each curve band by band and rescaling by its own level maps the picture into
reliability-diagram coordinates:
\begin{equation}\label{eq:reliability}
\bigl(\hat p(D\mid x),\,\E[Y\mid x]\bigr)
=\bigl(\slope_{\text{model}}(x)\cdot\hat p_D,\ \slope_{\text{emp}}(x)\cdot\pD\bigr),
\end{equation}
a point on the reliability diagram; perfect calibration places every band on the $45^\circ$
line. The two views carry identical information: the reliability diagram is the derivative of
the CAP comparison, and the CAP comparison is the cumulative integral of the reliability
diagram. The signed gap between the empirical and model-implied Ginis is a \emph{directional}
summary: its sign is the diagnostic. Model-implied below empirical signals compression, above
signals overconfidence (made precise in the example below). Its magnitude can cancel, since two
slope profiles that deviate in opposite directions integrate to the same area; for a
non-canceling magnitude we use the integrated calibration error (ICE), the absolute area
between the two CAP curves,
\begin{equation}\label{eq:ice}
\mathrm{ICE}=\int_0^1 \bigl|F_{\text{model}}(x\mid D)-F_{\text{emp}}(x\mid D)\bigr|\,dx,
\end{equation}
the cumulative form of the band-level calibration error and the CAP-coordinate analogue of the
expected calibration error (ECE; Guo et al., 2017). The calibration reference in CAP
coordinates is the \emph{empirical} curve itself (it plays the role of the $45^\circ$
diagonal in the reliability diagram), not the perfect-model boundary through $(\pD,1)$,
which is a ranking ideal irrelevant to calibration.

\begin{figure}[htbp]
\centering
\begin{tikzpicture}
\begin{axis}[
  width=0.49\textwidth,height=6.4cm,
  xmin=0,xmax=1,ymin=0,ymax=1,
  xlabel={$F(x)$}, ylabel={defaults / claims captured},
  legend style={at={(0.98,0.02)},anchor=south east,draw=none,font=\footnotesize},
  tick label style={font=\footnotesize}, label style={font=\footnotesize},
  grid=major, grid style={gray!18},
]
\addplot[capgray,dotted,thick] coordinates {(0,0)(1,1)};
\addplot[capblue,very thick,mark=*,mark size=1.4pt,name path=emp]
  coordinates {(0,0)(0.0870,0.25)(0.2609,0.50)(0.4783,0.75)(0.7913,0.95)(1,1)};
\addplot[capred,very thick,dashed,mark=square*,mark size=1.4pt,name path=mod]
  coordinates {(0,0)(0.0870,0.1757)(0.2609,0.4011)(0.4783,0.6483)(0.7913,0.9026)(1,1)};
\addplot[capred!12] fill between[of=emp and mod];
\legend{random,empirical CAP,model-implied CAP}
\end{axis}
\end{tikzpicture}\hfill
\begin{tikzpicture}
\begin{axis}[
  width=0.49\textwidth,height=6.4cm,
  xmin=0,xmax=0.55,ymin=0,ymax=0.55,
  xlabel={claimed $\hat p(D\mid x)$}, ylabel={observed $\E[Y\mid x]$},
  legend style={at={(0.02,0.98)},anchor=north west,draw=none,font=\footnotesize},
  tick label style={font=\footnotesize}, label style={font=\footnotesize},
  grid=major, grid style={gray!18},
]
\addplot[capgray,dotted,thick] coordinates {(0,0)(0.55,0.55)};
\addplot[only marks,capred,mark=*,mark size=2pt]
  coordinates {(0.0927,0.05)(0.1614,0.1111)(0.2259,0.20)(0.2575,0.25)(0.4013,0.50)};
\legend{$45^\circ$ (calibrated),bands A--E}
\end{axis}
\end{tikzpicture}
\caption{\textbf{Left:} empirical vs.\ model-implied CAP; the shaded area is the shape part of
the calibration error. \textbf{Right:} the same information as a reliability diagram, the
band-by-band derivative of the left panel, rescaled by each level
(Eq.~\ref{eq:reliability}). Band E sits \emph{above} the $45^\circ$ line (the model
under-predicts the riskiest band), while bands A--D sit below it, over-predicting: the
compressed slope profile is too timid at both ends.}
\label{fig:calib}
\end{figure}
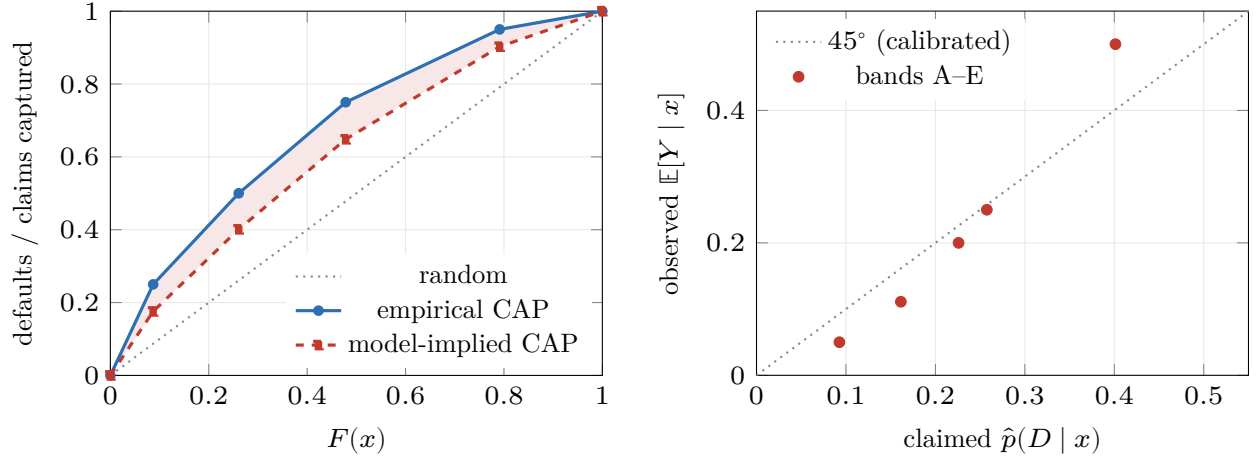

\paragraph{Comparison on the running example.} Suppose a model ranks the five bands correctly
but compresses risk in log-odds, $\operatorname{logit}\hat p(D\mid x)=\alpha+\beta\,
\operatorname{logit}p(D\mid x)$ with $(\alpha,\beta)=(-0.4,0.6)$, the standard
linear-in-log-odds recalibration form. The claimed PDs and the
resulting model-implied CAP are in
Table~\ref{tab:calib}; the two CAPs and the reliability diagram are in
Figure~\ref{fig:calib}. The level is nearly right ($\hat p_D=0.199$ vs.\ $\pD=0.174$) but the
shape is flattened, and the model-implied Gini collapses to $0.282$ against the empirical
$0.4395$. The sign of the gap is diagnostic: model-implied \emph{below} empirical means the
claims are less dispersed than the outcomes, a recalibration slope above one (the forward
compression $\beta=0.6$ inverts to a calibration slope $\approx 1.67$), while the reverse ordering,
model-implied above empirical, is the overconfidence signature. This level-versus-shape split is
the cumulative-coordinate counterpart of the calibration-curve factor decomposition of Voloshyn
and Voloshyn (2023), whose curve is linear in PD ($p(D\mid x)=ax+b$) rather than log-odds.

\begin{table}[htbp]
\centering
\small
\begin{tabular}{l S[table-format=1.4] S[table-format=1.4] S[table-format=1.4] S[table-format=1.4]}
\toprule
Sorted worst-first & {$\hat p(D\mid x)$} & {$\E[Y\mid x]$} & {$F_{\text{model}}(x\mid D)$} & {$F_{\text{emp}}(x\mid D)$}\\
\midrule
E & 0.4013 & 0.5000 & 0.1757 & 0.2500 \\
D & 0.2575 & 0.2500 & 0.4011 & 0.5000 \\
C & 0.2259 & 0.2000 & 0.6483 & 0.7500 \\
B & 0.1614 & 0.1111 & 0.9026 & 0.9500 \\
A & 0.0927 & 0.0500 & 1.0000 & 1.0000 \\
\midrule
\multicolumn{3}{l}{Gini (empirical $=0.4395$, model $=0.2823$)} & & \\
\bottomrule
\end{tabular}
\caption{Comparison mode on the running example. The model preserves the ranking but
compresses the slope profile; the gap between the empirical and model-implied CAPs is the
calibration error.}
\label{tab:calib}
\end{table}

\begin{figure}[htbp]
\centering
\begin{tikzpicture}
\begin{axis}[
  width=0.49\textwidth,height=6.4cm, xmin=0,xmax=1,ymin=0,ymax=1.02,
  xlabel={$F(x)$ (worst first)}, ylabel={$F(x\mid D)$},
  tick label style={font=\footnotesize}, label style={font=\footnotesize},
  grid=major, grid style={gray!15},
  legend style={at={(0.97,0.03)},anchor=south east,draw=none,font=\scriptsize},
]
\addplot[capgray,dotted,thick] coordinates {(0,0)(1,1)}; \addlegendentry{random}
\addplot[black,very thick] table[x=Fx,y=Femp]{data/compare_cap.dat}; \addlegendentry{empirical}
\addplot[capblue,very thick,dashed] table[x=Fx,y=FmodA]{data/compare_cap.dat}; \addlegendentry{Model A claims}
\addplot[capred,very thick,dashed] table[x=Fx,y=FmodB]{data/compare_cap.dat}; \addlegendentry{Model B claims}
\node[font=\scriptsize,anchor=north west,align=left] at (axis cs:0.05,0.99)
 {Gini $=0.59$ (both)\\[1pt]\textcolor{capblue}{A claims $0.58$}\\\textcolor{capred}{B claims $0.78$}};
\end{axis}
\end{tikzpicture}\hfill
\begin{tikzpicture}
\begin{axis}[
  width=0.49\textwidth,height=6.4cm, xmin=0,xmax=0.85,ymin=0,ymax=0.85,
  xlabel={predicted $\hat p(D\mid x)$}, ylabel={observed $\E[Y\mid x]$},
  tick label style={font=\footnotesize}, label style={font=\footnotesize},
  grid=major, grid style={gray!15},
  legend style={at={(0.03,0.97)},anchor=north west,draw=none,font=\scriptsize},
]
\addplot[capgray,dotted,thick] coordinates {(0,0)(0.85,0.85)}; \addlegendentry{calibrated}
\addplot[capblue,thick,mark=*,mark size=1.4pt] table[x=pA,y=EA]{data/compare_rel.dat}; \addlegendentry{Model A}
\addplot[capred,thick,mark=square*,mark size=1.4pt] table[x=pB,y=EB]{data/compare_rel.dat}; \addlegendentry{Model B (overconfident)}
\end{axis}
\end{tikzpicture}
\caption{Two models with identical ranking, hence identical empirical Gini (left), but different
calibration (right). Both share the empirical CAP; Model~A's claimed CAP sits on it (claimed
Gini $0.58$ vs.\ empirical $0.59$), while the overconfident Model~B's claimed CAP overstates
discrimination (claimed Gini $0.78$) and its reliability points bow off the $45^\circ$ line,
a gap the area-based Gini cannot see.}
\label{fig:twomodels}
\end{figure}
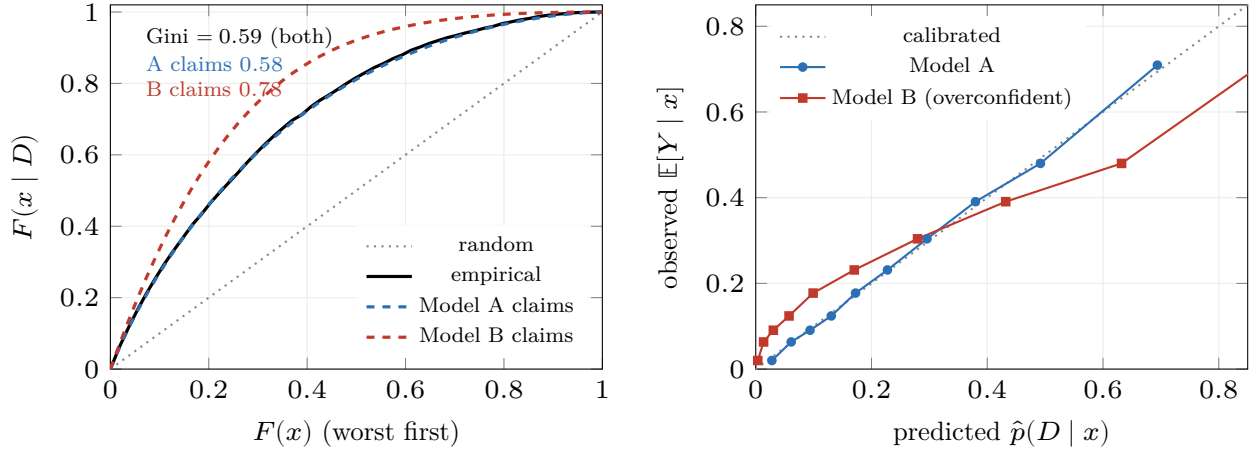

\paragraph{Two models, one Gini.} The diagnostic earns its keep when two models rank borrowers
identically but disagree on the probabilities, a difference the Gini cannot see.
Figure~\ref{fig:twomodels} puts two such models on one portfolio: their scores are monotone
transforms of each other, so they share a single empirical CAP and a single Gini ($0.59$).
Model~A claims the realized PDs; Model~B is overconfident, pushing its claims too far in
log-odds. The area cannot separate them (the empirical curve is the same for both), yet
their model-implied curves diverge: A's claimed Gini ($0.58$) matches the empirical, while B's
($0.78$) overstates its own discrimination, and B's reliability points bow off the $45^\circ$
line. The gap between each model's implied Gini and the empirical Gini is the calibration
summary useful for practitioners.

\section{The continuous case: density estimation}\label{sec:continuous}

The discrete derivation transfers verbatim to continuous scores once the three distributions
are estimated as densities rather than band frequencies. The CAP slope \eqref{eq:cap-slope} is
a ratio of densities,
\begin{equation}
\frac{dF(x\mid D)}{dF(x)}=\frac{f(x\mid D)}{f(x)}=\frac{\PDx}{\pD},
\end{equation}
and the weight of evidence \eqref{eq:woe-def} is a log-ratio of class-conditional densities,
$\WOE(x)=\ln f(x\mid D)-\ln f(x\mid\bar D)$. Both reduce to estimating the class-conditional
densities $f(x\mid D)$ and $f(x\mid\bar D)$. Alvarez-Melis (2019, 2021) does exactly this for
weight of evidence explanations: fit a kernel density estimate $\hat f(x\mid c)$ (or a Gaussian
model) to each class's scores and take the log-density difference $\widehat{\WOE}(x)=\ln\hat
f(x\mid D)-\ln\hat f(x\mid\bar D)$, the continuous limit of the band-counting WOE in
Section~\ref{sec:woe}. The same
estimates feed the standardized PD $\hat f(x\mid D)/\hat f(x)$ and, integrated, the CAP itself.
In the continuous reading the rating bands of Section~\ref{sec:setup} are a histogram
discretization of these densities, and the monotone PD scale is the requirement that the
density ratio $f(x\mid D)/f(x\mid\bar D)$ be monotone in the score.

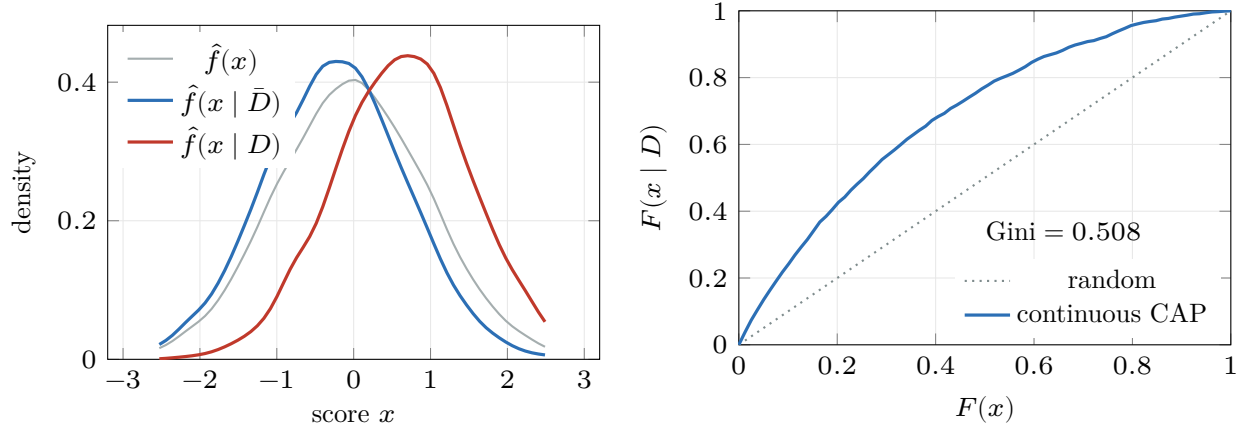
\begin{figure}[htbp]
\centering
\begin{tikzpicture}
\begin{axis}[
  width=0.49\textwidth,height=6cm,
  xlabel={score $x$}, ylabel={density},
  xmin=-3.2,xmax=3.2, ymin=0,
  legend style={at={(0.02,0.98)},anchor=north west,draw=none,font=\footnotesize},
  tick label style={font=\footnotesize}, label style={font=\footnotesize},
  grid=major, grid style={gray!18},
]
\addplot[capgray!70,thick] table[x=x,y=fX]{data/kde_densities.dat};
\addlegendentry{$\hat f(x)$}
\addplot[capblue,very thick] table[x=x,y=fG]{data/kde_densities.dat};
\addlegendentry{$\hat f(x\mid\bar D)$}
\addplot[capred,very thick] table[x=x,y=fD]{data/kde_densities.dat};
\addlegendentry{$\hat f(x\mid D)$}
\end{axis}
\end{tikzpicture}\hfill
\begin{tikzpicture}
\begin{axis}[
  width=0.49\textwidth,height=6cm,
  xlabel={$F(x)$}, ylabel={$F(x\mid D)$},
  xmin=0,xmax=1,ymin=0,ymax=1,
  legend style={at={(0.98,0.02)},anchor=south east,draw=none,font=\footnotesize},
  tick label style={font=\footnotesize}, label style={font=\footnotesize},
  grid=major, grid style={gray!18},
]
\addplot[capgray,dotted,thick] coordinates {(0,0)(1,1)};
\addlegendentry{random}
\addplot[capblue,very thick] table[x=Fx,y=FxD]{data/kde_cap.dat};
\addlegendentry{continuous CAP}
\node[font=\footnotesize] at (axis cs:0.66,0.34) {$\mathrm{Gini}=0.508$};
\end{axis}
\end{tikzpicture}
\caption{\textbf{Left:} the three kernel density estimates. Bayes reweights $\hat f(x)$ toward
high scores to give the defaulter density $\hat f(x\mid D)$. \textbf{Right:} integrating
$\hat f(x\mid D)$ against $\hat f(x)$ gives the continuous CAP; its area returns the same Gini
as the pair-counting route.}
\label{fig:kde-dens}
\end{figure}

\paragraph{A worked KDE example.} As an illustration (a numerical check that the identity
survives the continuous limit, not a proposal for a production WOE estimator), take $n=8000$
borrowers with a continuous score $x\sim\mathcal N(0,1)$ and a logistic truth
$\operatorname{logit}p(D\mid x)=\beta_0+\beta_1 x$, $\beta_0=-1.3$, $\beta_1=1.1$
($\pD=0.258$). We fit three Gaussian kernel density estimates ($\hat f(x\mid D)$ on the
defaulters' scores, $\hat f(x\mid\bar D)$ on the non-defaulters', $\hat f(x)$ on all scores, each with
the default Scott's-rule bandwidth; Figure~\ref{fig:kde-dens}, left) and read the paper's
quantities off the densities, using none of the ground truth.

\begin{figure}[htbp]
\centering
\begin{tikzpicture}
\begin{axis}[
  width=0.49\textwidth,height=6cm,
  xlabel={score $x$}, ylabel={$\WOE(x)$},
  xmin=-3.2,xmax=3.2,
  legend style={at={(0.02,0.98)},anchor=north west,draw=none,font=\footnotesize},
  tick label style={font=\footnotesize}, label style={font=\footnotesize},
  grid=major, grid style={gray!18},
]
\addplot[capred,only marks,mark=*,mark size=0.7pt] table[x=x,y=woe_kde]{data/kde_woe.dat};
\addlegendentry{$\ln\hat f(x\mid D)-\ln\hat f(x\mid\bar D)$}
\addplot[capgreen,very thick,dashed] table[x=x,y=woe_true]{data/kde_woe.dat};
\addlegendentry{true (slope $\beta_1{=}1.1$)}
\end{axis}
\end{tikzpicture}\hfill
\begin{tikzpicture}
\begin{axis}[
  width=0.49\textwidth,height=6cm,
  xlabel={score $x$}, ylabel={$p(D\mid x)$},
  xmin=-3.2,xmax=3.2, ymin=0,ymax=1,
  legend style={at={(0.02,0.98)},anchor=north west,draw=none,font=\footnotesize},
  tick label style={font=\footnotesize}, label style={font=\footnotesize},
  grid=major, grid style={gray!18},
]
\addplot[capred,only marks,mark=*,mark size=0.7pt] table[x=x,y=phat]{data/kde_pd.dat};
\addlegendentry{$\pD\,\hat f(x\mid D)/\hat f(x)$}
\addplot[capgreen,very thick,dashed] table[x=x,y=ptrue]{data/kde_pd.dat};
\addlegendentry{true $\sigma(\beta_0+\beta_1 x)$}
\end{axis}
\end{tikzpicture}
\caption{Illustration that the identity survives the continuous limit (single draw, Scott's-rule
bandwidth). \textbf{Left:} the KDE weight of evidence (points) is linear in the score; fitting a
line gives a logit slope of $1.04$ vs.\ true $1.1$, the gap being kernel-smoothing attenuation
(Section~\ref{sec:continuous}). \textbf{Right:} rescaling the densities by the prior reproduces
the logistic PD curve. Code: \texttt{scripts/continuous\_kde.py}.}
\label{fig:kde-rec}
\end{figure}
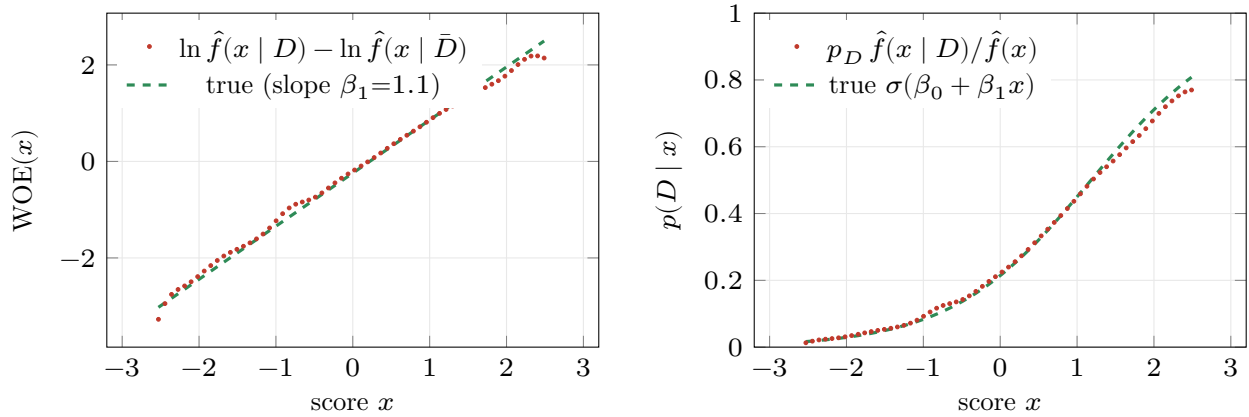

The density route reproduces the calibration curve it never saw. The estimated weight of
evidence $\ln\hat f(x\mid D)-\ln\hat f(x\mid\bar D)$ is, by \eqref{eq:woe-def}, a straight line
in the score whenever the ground truth is logistic; on this draw, fitting a line to the KDE-WOE over
the bulk of the data gives a logit slope of $1.04$ against the true $\beta_1=1.1$ and an
intercept of $-0.22$ against $\beta_0-\operatorname{logit}\pD=-0.24$ (Figure~\ref{fig:kde-rec},
left). The residual gap is structural, not sampling noise: kernel smoothing convolves both
class densities and attenuates the log-density-difference slope toward zero, a bias that shrinks with the
bandwidth and vanishes as $h\to0$. (We report a single draw without standard errors; the point
is the limiting identity, not a calibrated estimator.) Rescaling the same densities by the
prior, $\hat p(D\mid x)=\pD\,\hat f(x\mid D)/\hat f(x)$, reproduces the logistic calibration
curve (Figure~\ref{fig:kde-rec}, right). The CAP of the continuous scores
(Figure~\ref{fig:kde-dens}, right) returns $\mathrm{Gini}=0.508$ identically from the CAP-area
route $(2A-1)/(1-\pD)$ and the pair-counting route $2\,\mathrm{AUC}-1$, both read off the
same empirical CDF, so the agreement is the exact rank identity of Section~\ref{sec:somers}.
The discrete five-band example and the continuous density estimate have the same identity.

\section{Relation to prior work}\label{sec:related}

The slope identity originates with Falkenstein, Boral and Carty (2000) as a discrete
relationship between the default-frequency and power curves, stated without derivation and
without reference to Bayes. Van der Burgt (2008) turns it into a calibration method for
low-default portfolios by fitting a parametric CAP and differentiating; Tasche (2009) studies
the statistical behavior of that method; van der Burgt (2019) writes the continuous identity
$\PDx=\pD\,dy/dx$, derives the per-grade version from Bayes' law, and builds the rating scale
by partitioning the CAP, finding the grade count to scale as a power law of the accuracy ratio
and $\pD$. Voloshyn and Voloshyn (2023) substitute Bayes' theorem into the area integral to
write the Gini as a functional of the calibration curve and decompose Gini changes into cutoff,
population, payment-discipline, and discriminatory-power factors.

Against this lineage, the present paper (i) derives the slope identity as Bayes' theorem in
cumulative coordinates, unifying the discrete per-band Bayes step and the continuous
derivative; (ii) carries the derivation through the odds form to place weight of evidence, the
Bayes factor, and information value inside the same CAP geometry, connecting to the
information-theoretic scorecard framework of Sudjianto and Burakov (2025) and the
weight of evidence explanations of Alvarez-Melis (2019, 2021); and (iii) shows that the
accuracy ratio, Somers' $D_{xy}$, and the Gini are one number computed three ways, and
uses the identity in comparison mode as a calibration diagnostic equivalent to the reliability
diagram. The exact $\AR=(2A-1)/(1-\pD)$, required for the Somers' $D_{xy}$ identity, follows
Thomas (2009) and matches van der Burgt (2019); the $1-\pD$ correction is what the
$2A-1$ approximation discards.

\section{Conclusion}

The cumulative accuracy profile is usually read as a power object, and its area as a
power-only summary. Read through Bayes' theorem, its slope is a calibration object: the slope
is the posterior rescaled by the prior, the log-ratio of the bad and good slopes is the weight
of evidence, the divergence of the two class-conditional densities is the information value, and the area,
corrected by $1-\pD$, is at once the accuracy ratio, Somers' $D_{xy}$, and the ROC Gini. Run
in comparison mode, the same identity makes the CAP a reliability diagram in cumulative
coordinates. What the identity cannot do, absent a parametric assumption, is invert a single
area into a full calibration curve: slope times level is identity, but the shape of the
slope profile is assumption. Future work includes standard errors for the standardized PD and the Gini gap under the
pair-counting view, and a calibration threshold on the integrated calibration error (ICE) of
Eq.~\eqref{eq:ice}.

\appendix
\section{Code and data availability}\label{app:code}
Every number, table, and figure in this paper is reproduced by the code at
\begin{center}
\url{https://github.com/deburky/gini-bayes-paper}
\end{center}
(\texttt{numpy}, plus SciPy for the kernel density estimates; run with \texttt{uv run python}).
The discrete five-band example of Sections~\ref{sec:setup}--\ref{sec:calib} (the Bayes
reweighting, the accuracy ratio computed three ways, the weight of evidence and information
value, and the comparison-mode calibration gap) is in \texttt{scripts/five\_band\_example.py};
the continuous kernel-density example of Section~\ref{sec:continuous} is in
\texttt{scripts/continuous\_kde.py}, which also writes the figures' data files. The five-band
quantities use the exact rational values of Table~\ref{tab:data}; the comparison-mode model is
the linear-in-log-odds recalibration of Section~\ref{sec:calib}.


\begin{thebibliography}{99}
\setlength{\itemsep}{0.3em}

\bibitem{alvarezmelis2019} Alvarez-Melis, D., Jaakkola, T., and Jegelka, S. (2019).
\emph{Weight of Evidence as a Basis for Human-Oriented Explanations}.
arXiv:1910.13503.

\bibitem{alvarezmelis2021} Alvarez-Melis, D., Kaur, H., Daum\'e III, H., Wallach, H., and
Wortman Vaughan, J. (2021). \emph{From Human Explanation to Model Interpretability: A
Framework Based on Weight of Evidence}. arXiv:2104.13299.

\bibitem{falkenstein2000} Falkenstein, E., Boral, A., and Carty, L.~V. (2000).
\emph{RiskCalc for Private Companies: Moody's Default Model}. Rating Methodology, Global
Credit Research, Moody's Investors Service.

\bibitem{good1950} Good, I.~J. (1950). \emph{Probability and the Weighing of Evidence}.
Charles Griffin, London.

\bibitem{good1985} Good, I.~J. (1985). Weight of evidence: a brief survey. In J.~M. Bernardo,
M.~H. DeGroot, D.~V. Lindley, and A.~F.~M. Smith (eds.), \emph{Bayesian Statistics 2},
pp.~249--270. North-Holland.

\bibitem{guo2017} Guo, C., Pleiss, G., Sun, Y., and Weinberger, K.~Q. (2017). On calibration
of modern neural networks. In \emph{Proceedings of the 34th International Conference on Machine
Learning (ICML)}, pp.~1321--1330.

\bibitem{newson2002} Newson, R. (2002). Parameters behind ``nonparametric'' statistics:
Kendall's $\tau$, Somers' $D$ and median differences. \emph{The Stata Journal} 2(1), 45--64.

\bibitem{somers1962} Somers, R.~H. (1962). A new asymmetric measure of association for
ordinal variables. \emph{American Sociological Review} 27(6), 799--811.

\bibitem{sudjianto2025} Sudjianto, A., and Burakov, D. (2025). \emph{An Information-Theoretic
Framework for Credit Risk Modeling: Unifying Industry Practice with Statistical Theory for
Fair and Interpretable Scorecards}. arXiv:2509.09855.

\bibitem{tasche2005} Tasche, D. (2005). Rating and probability of default validation. Working
Paper 14, Studies on the Validation of Internal Rating Systems, Basel Committee on Banking
Supervision, Bank for International Settlements.

\bibitem{tasche2009} Tasche, D. (2009). \emph{Estimating Discriminatory Power and PD Curves
When the Number of Defaults Is Small}. arXiv:0905.3928.

\bibitem{thomas2009} Thomas, L.~C. (2009). \emph{Consumer Credit Models: Pricing, Profit and
Portfolios}. Oxford University Press.

\bibitem{vdb2008} van der Burgt, M.~J. (2008). Calibrating low-default portfolios, using the
cumulative accuracy profile. \emph{Journal of Risk Model Validation} 1(4), 17--33.

\bibitem{vdb2019} van der Burgt, M.~J. (2019). Calibration and mapping of credit scores by
riding the cumulative accuracy profile. \emph{Journal of Credit Risk} 15(1), 1--25.
\url{https://doi.org/10.21314/JCR.2018.240}.

\bibitem{vdb2019talk} van der Burgt, M.~J. (2019). \emph{Calibration and Mapping of Credit
Scores by Riding the Cumulative Accuracy Profile}. Presentation, EuroBanking 2019, Ljubljana,
26--29 May (slide~3, ``CAP as calibration tool'').

\bibitem{voloshyn2023} Voloshyn, M., and Voloshyn, I. (2023). \emph{On Factors Affecting the
Change in the Gini Coefficient of the Credit Scoring Model}. ScienceOpen Preprint.
\url{https://doi.org/10.14293/PR2199.000388.v1}.

\end{thebibliography}
\end{document}